\documentclass[aps,pre,twocolumn,superscriptaddress,showpacs,showkeys ]{revtex4-2}

\usepackage{graphicx}                   
\usepackage{hyperref}                   
\usepackage{amsmath,amssymb}            
\usepackage{epstopdf}
\usepackage{subcaption}					
\usepackage{gensymb}
\usepackage{siunitx}

\usepackage{color}
\definecolor{orange}{rgb}{1,0.5,0}
\definecolor{brown}{rgb}{0.65, 0.16, 0.16}
\definecolor{phlox}{rgb}{0.87, 0.0, 1.0}

\graphicspath{{figs/}}					

\begin{document}
	
	\title{Multifractal analysis of Earthquakes in Central Alborz, Iran; A phenomenological self-organized critical Model}
	
	\author{M. Rahimi-Majd}
	\affiliation{Department of Physics, Shahid Beheshti University, 1983969411, Tehran, Iran}
	
	\author{T. Shirzad}
	\affiliation{Institute of Geophysics, Polish Academy of Sciences - 01-452, Warsaw, Poland}
	
	\author{M. N. Najafi}
	\affiliation{Department of Physics, University of Mohaghegh Ardabili, P.O. Box 179, Ardabil, Iran}
	\email{morteza.nattagh@gmail.com}
	
	\begin{abstract}
		This paper is devoted to a phenomenological study of the earthquakes in central Alborz, Iran. Using three observational quantities, namely weight function, quality factor, and velocity model in this region, we develop a phenomenological dissipative sandpile-like model which captures the main features of the system, especially the average activity field over the region of study. The model is based on external stimuli, the location of which are chosen (\textbf{I}) randomly, (\textbf{II}) on the faults, (\textbf{III}) on the highly active points in the region. We analyze all these cases and show some universal behaviors of the system depending slightly on the method of external stimuli. The multi-fractal analysis is exploited to extract the spectrum of the Hurst exponent of time series obtained by each of these schemes. Although the average Hurst exponent depends on the method of stimuli (the three cases mentioned above), we numerically show that in all cases it is lower than $0.5$, reflecting the anti-correlated nature of the system. The lowest average Hurst exponent is for the case (\textbf{III}), in such a way that the more active the stimulated sites are the lower the value for the average Hurst exponent is obtained, i.e. the larger earthquakes are more anticorrelated. However, the different activity fields in this study provide the depth of the basement, the depth variation (topography) of the basement, and an area that can be the location of the future probability event. 
	\end{abstract}
	
	
	\maketitle
	
	\section{Introduction}
	
	It is widely believed that the earthquake is a self-organized critical system. When and earthquake occurs and a fault slips, it causes an excess in the tension of the neighboring regions. The motion of the neighboring regions depends on their local strain, so that if it exceeds a threshold, then it slips. This dynamics is reminiscent of the sandpile dynamics, first invented by Bak-Tang-Wiesenfeld (BTW) model, where the energy spreads the system based on similar dynamic rules: when the local energy exceeds a threshold, the site topples, rising the energy of the neighboring sites by one unit. BTW and some other variants~\cite{paczuski1996universality,lise2002nonconservative} are too unrealistic to give an acceptable description of the earthquake, e.g. it cannot explain $\frac{1}{f}$ noise that is seen in reality~\cite{moghadam2018power}. For more realistic situations, one needs more detailed model paying attention to the structure of the earth and also the information on how the seismic activities affect each other, which translates to how a perturbation propagates from region to region. The latter is crucial and pretty complicated since it depends on the material content of the Earth's interior where the signal propagates. Recently, using a virtual seismometers, the correlations between seismic activities were incorporated in the model, based on which a complex network was designed on top of which sandpile dynamics were implemented~\cite{najafi2020avalanches}. The virtual seismometer can provide inter-event empirical Green's function in the Earth's interior. Therefore, using unconventional form of seismic interferometry, one earthquake beneath the Earth's surface can turn to a receiver where recorded another event waveform~\cite{curtis2009virtual}. One may think of this problem from another point of view: the cross-correlation between previously happened earthquakes gives us a set of valuable information about the structure of the earth and the signal propagation in the region. For example, suppose that we have a map (the place and the magnitude, \textbf{M}) of previous earthquakes in the region of interest, along with the corresponding time series. Cross-correlation between the events gives us criteria of how events are related and to which extend they are correlated. \\
	
	As the case study, we focus on the earthquakes in central Alborz in the present paper. Alborz range with seismic active east-west trending mountain belt extends across the north of Iran. The south Caspian block to the north, central Iran micro-plateau to the south surround the central Alborz with several folds and various faults (see FIG.~\ref{fig:Phenomenona}). Three major tectonic events can invoke for Central Alborz including: \textit{I}) shortening (led to thrusting and folding~\cite{guest2006late}), \textit{II}) extension (led to Damavand Volcanism~\cite{davidson2004geology}), and \textit{III}) collision-related compression (from middle Miocene to recent~\cite{guest2007late}). Studies of the crustal deformation by GPS measurements ~\cite{vernant2004present} indicated compression between the Central Iranian micro-plateau and South Caspian blocks evince a partitioning ~5 $mmyr^{-1}$ range-perpendicular and ~4 $mmyr^{-1}$ along range-parallel. This region is very active and experience many large, and catastrophic earthquakes (e.g., the catastrophic Manjil-Rudbar in July $20^{th}$, 1990 with \textbf{M}$_W$ $7.4$), which are associated with major active faults (red line in FIG.~\ref{fig:Phenomenona}).
	Historical~\cite{ambraseys1982history,berberian1999patterns,berberian2001contribution,berberian2017tehran} and instrumental recorded earthquakes in the central Alborz represent that many faults have the potential of an earthquake up to \textbf{M} $7.5$.
	
	The character of the seismic wave propagation effects directly depends on the nature of the distribution of the elastic parameters within the Earth's interior. Other than the seismic source functions (focal mechanisms, the rupturing algorithms, and time duration, etc.), nowadays, these elastic parameters of the seismic wave propagation (e.g., seismic velocity structures, attenuation models, etc.) are calculated using these recorded waveforms with a combination of the classical and new seismological processing methods. Today, different methods exist for calculating 1D, 2D, or 3D structures of Earth and various properties of seismic records may be used, including amplitudes, travel times, full waveforms, etc. Inversion of the arrival times of seismic (body or surface) waves is one of the routines and popular techniques for imaging Earth's interior~\cite{kennett1988subspace,barmin2001fast}. Several studies about the crustal velocity structure of the Alborz range have been done using the traveltime tomography method and different kinds of seismic recording components. The 1D velocity model with two sedimentary layers ($V_P \leq 6.0 $ $ kms^{-1}$; with a total thickness of 8 km) overlying two crystalline layers ($6.0 < V_P \leq 6.3 kms^{-1}$) was obtained by Ashtari et al.~\cite{ashtari2005microseismicity} which employed the first arrival P- and S-waves inversion. This 1D model was updated by ~\cite{abbassi2010crustal} with a similar method (first arrival time of P- and S-waves) by deploying dense temporary seismic stations on the southern edge of Central Alborz. Recently, the 1D model was updated by ~\cite{soltanimoghadam2019improved} by combining recorded data of all available run temporary and permanent seismic networks.\\
	
	The 3D first arrival P-wave velocity model for Alborz Mountains was calculated by ~\cite{maheri2020upper,rezaeifar2020regional}.
	Although previous surface wave tomographic models (e.g.,  teleseismic model ~\cite{shad2011new}; ambient seismic noise results ~\cite{mottaghi2013ambient,kaviani2020crustal}) can address relatively good resolution on a regional scale, they do not have any resolution power to provide an insight into the crustal structure (up to 30 km) in Central Alborz. Several P-wave tomographies (e.g. the study of Tehran City by Shirzad et al.~\cite{shirzad2018shallow}), surface wave tomography (e.g. Tehran City by ~\cite{shirzad2018shallow}, North Tehran and Mosha fault junction by ~\cite{shirzad2018shallow}), and radial anisotropy (e.g. Tehran city by ~\cite{shirzad2014shallow} and North Tehran and Mosha fault junctions by ~\cite{naghavi2019radial}) were studied which provide the velocity model for a local and/or for a part of Central Alborz. Moreover, seismic attenuation, which generally leads to the decrease in the amplitude of seismic waves, has been studied by several researchers~\cite{rahimi2010variation,naghavi2012lg,farrokhi2015estimation} in central Alborz.\\
	
	In this paper, we use the already established data on the quality factor, velocity model, and cross-correlations of seismic activity of the region and calculate the weight field over the system under study. Then we apply a dissipative avalanche dynamics to the system by designing an activity-propagation algorithm based on the phenomenological parameters that were explored.\\
	
	The paper has been organized as follows: in the next section, we present the previous analysis on the region under study and introduce the phenomenological quantities of interest. In Sec.~\ref{SEC:Weight}, we introduce our model for the weight function, based on which our self-organized critical model is defined. We describe our model in Sec.~\ref{SEC:Model}, where the phenomenological parameters of the previous section are employed. Section~\ref{NUMDet} is devoted to the numerical results and the activity field. The multi-fractal analysis is presented in Sec.~\ref{SEC:MFA} where the generalized variance of the activity time series is analyzed, and the corresponding Hurst exponent is extracted. We close the paper with a conclusion. 
	
	\section{Observational data, Our phenomenological model}
	{A network with evenly grid space was used in this study, which has been achieved by processing the waveforms of the occurred earthquakes in the Central Alborz region. Calculating the weight function between cells involves the interferometry of the recorded waveforms, attenuation, and seismic velocity model of the study area.
		
		\subsection{Earthquake Dataset}
		We processed all earthquakes that occurred in the Central Alborz with \textbf{M} $\geq 2.5$ between 2006 and 2021. These events were recorded by 48 seismic stations operated by three permanent seismic networks that included the following: (\textit{I}) \textbf{Ir}anian \textbf{S}eismological \textbf{C}enter, IrSC, (\textit{II}) \textbf{I}nterational \textbf{I}nstitute of \textbf{E}arthquake \textbf{E}ngineering and \textbf{S}eismology, IIEES, and (\textit{III}) \textbf{T}ehran \textbf{D}isaster \textbf{M}anagement and \textbf{M}itigation \textbf{O}rganization, TDMMO. Seismic stations of IrSC and TDMMO have been equipped with SS1 (with 50 sps), CK1 (with 71.43 sps) short period, while the IIEES has been supported by Guralp CMG-3 broadband with 100 sps, respectively. From more than 3000 events (see FIG.~\ref{fig:Phenomenona}a), we selected 372 earthquakes based on following criteria: earthquakes magnitudes \textbf{M} $\geq$ 4, both horizontal and vertical location uncertainties less than 2 km, \textit{RMS} $\leq $0.2 s, azimuthal gap $\leq$ 180$\degree$, and event recorded at least by 10 seismic stations.
		
		\begin{figure*}
			\centerline{\includegraphics[scale=0.31]{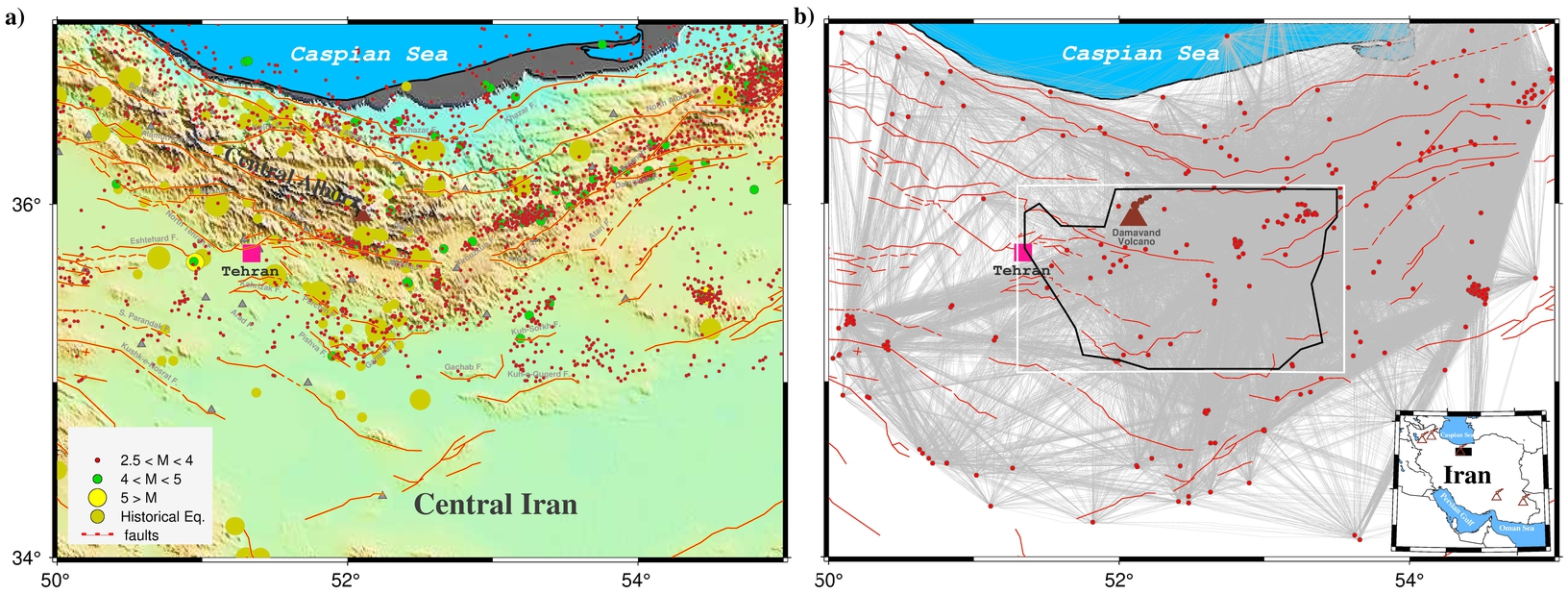}}
			\caption{The map of the study area. The know faults, circles and volcano depict by solid red lines, earthquakes and brown triangle, respectively. The pink square represent Tehran, capital of Iran. The historical events reported by ~\cite{ambraseys1982history,berberian1999patterns,berberian2001contribution,berberian2017tehran} and instrumental earthquakes in the central Aborz located by \textbf{Ir}anian \textbf{S}eismological \textbf{C}enter, IrSC \protect\footnote{\url{http://irsc.ut.ac.ir}}. The white box shows a region with a fair resolution (FIGs. 2-6). The black rectangular in the inset map indicates the study area.}
			\label{fig:Phenomenona}
		\end{figure*}
		
		\subsection{Weight Models}\label{SEC:Weight}
		{The size and strength of heterogeneities in the Earth's interior can affect the coupling weight of neighborhood cells. The event interferometry approach~\cite{curtis2009virtual} can provide the possibility of retrieving the coupling weight without a set of dense stations and/or expensive seismic imaging. For each event, we selected waveforms of the vertical component (\textit{Z}-component) with signal-to-noise ratio, \textit{SNR} $\geq$ 4.0, epicentral distance, 15 km $\leq$ \textit{$dist$} $\leq$ 180 km, and then single station data preparation was done. This preparation includes removing mean, trend, correcting of instrument response, pre-filtering (with a 5-point zero-phase bandpass Butterworth filter) in a period range of 1 to 30 s, and then running time (one-bit) and frequency (whitening) domain normalizations to suppress the influence of instrument irregularities, human activities, and source time functions with nonuniform energy. The maximum amplitude of the envelope function within the expected signal window (1.5 - 3.5 km/s) to the root-mean-square, \textit{rms}, is the \textit{SNR} definition in this study~\cite{pedersen2007influence}. The amplitudes of the waveform parts beyond the expected signal window were transferred to zero. Next, we took the waveform part from the origin time, $t_o$, to the end of the Rayleigh coda wave (1.5 km/s) for the signals-cross-correlation procedure. Then, we applied the signal-cross-correlation (hereafter \textit{SCC}) operator on the prepared waveforms of a pair-event which are recorded by a common station in the alignment of the inter-event line. The total number of inter-event raypaths depicts in FIG.~\ref{fig:Phenomenona}b which is 47,864 paths. The \textit{SCC} can be summarized
			\begin{widetext}
				\begin{equation}\label{eq:threshold}
					^{T}\textbf{m}_{1} ^{T}\textbf{m}_2\partial_1\partial_2\Gamma(r_2|r_1)= \int_{S} \{ \textbf{u}(r'|r_2).\textbf{T}^*(r'|r_2)-\textbf{T}(r'|r_1).\textbf{u}^*(r'|r_1) \} dr 
				\end{equation}
			\end{widetext}
			where $^T\textit{\textbf{m}}$, $\textit{$\partial$}$, \textbf{\textit{$\Gamma$}}, \textit{r}, \textit{\textbf{u}}, and \textit{\textbf{T}} are the moment tensor solution, spatial gradient, inter-event Green's function, coordinate vector, displacement and traction, respectively. Moreover, the indices \textit{1} , \textit{2}, and \textit{S} represent event1, event2, and Earth's surface, respectively. Because we took into account only vertical components of signals, the full moment tensor solutions reduce from nine components (\textit{$^{T}\textbf{m}_{RR}$}, \textit{$^{T}\textbf{m}_{RT}$}, \textit{$^{T}\textbf{m}_{RZ}$}, \textit{$^{T}\textbf{m}_{TR}$}, \textit{$^{T}\textbf{m}_{TT}$}, \textit{$^{T}\textbf{m}_{TZ}$}, \textit{$^{T}\textbf{m}_{ZR}$}, \textit{$^{T}\textbf{m}_{ZT}$}, \textit{$^{T}\textbf{m}_{ZZ}$}) to one component (\textit{$^{T}\textbf{m}_{ZZ}$}), which is a constant value. The effect of these constants (\textit{$^{T}\textbf{m}_{ZZ_1}$} and \textit{$^{T}\textbf{m}_{ZZ_2}$}) have been removed using a simple time-domain normalization operator in the data preparation step. Then, the maximum amplitude of the absolute \textit{SCC} function, $_{SCC}A_{max}$, was extracted and attributed to the corresponding cells along the inter-event raypath.
			
			Given the limitations and non-uniform distribution in raypath coverage, we used a grid base tomography procedure to obtain \textit{SCC} weight for all evenly grid space in the study area. For tomography, the observed data, \textit{$\textbf{d}^{obs}$}, is $t  _{_{SCC}A_{max}}$ (the arrival time of $_{SCC}A_{max}$) and model, $\textbf{m}^{\text{cal}}$, could be the $_{SCC}A_{max}$. Therefore, the relation between data and model is
			
			\begin{equation}\label{eq:threshold}
				\textbf{d}^{\text{obs}}-\textbf{d}^{\text{cal}}=\textbf{G}(\textbf{m}^{\text{true}})-\textbf{G}(\textbf{m}^{\text{est}})
			\end{equation}
			The calculated data, \textit{$d^{\text{cal}}$}, can be calculated using \textbf{F}ast \textbf{M}arching \textbf{M}ethod (FMM;~\cite{sethian1996fast,rawlinson2004wave}) grid base algorithm. When \textbf{G}, Green's function, is a linear or near-linear function this formula can be  \\
			\begin{equation}\label{eq:threshold}
				\delta \textbf{d}= \textbf{G} \delta \textbf{m} \Longleftrightarrow \delta \textbf{m}= \textbf{G}^{-1} \delta \textbf{d} 
			\end{equation}
			An iterative linearized damped-least squares inversion procedure can apply to minimize the observed and calculated data misfit~\cite{kennett1988subspace,rawlinson2005fmst}. Using Gauss-Newton gradient method, the relation (3) can be
			\begin{equation}\label{eq:delta_m}
				\delta \textbf{m}=[ \textbf{G}^{Tr} C_{d}^{-1} \textbf{G} + \epsilon C_{m}^{-1}+\eta D^{Tr} D] ^{-1} \textbf{G}^{Tr} C_{d}^{-1}\delta \textbf{d} 
			\end{equation}
			where $\epsilon$ and $\eta$ are the damping, and smoothing regularization parameters, and also $C_{d}^{-1}$ and $C_{m}^{-1}$ remark the data, and model errors, respectively, and \textit{$Tr$} represent the transpose operator. To solve equation~\ref{eq:delta_m}, the study area was divided by an even grid cell size of $14 km \times 14 km$, and the average of $\textit{A}_{\text{max}}$ applied as initial input model, $m_0$. Also, the regularization parameters, $\epsilon$, and $\eta$, were obtained by standard L-curve by considering a trade-off between data misfit and model roughness. To stabilize the result, we used those observed data, $\textit{A}_{\text{max}}$, with residuals less than two standard deviations (2$\sigma_{\textit{A}_{\text{max}}}$) in the inversion procedure. FIG.~\ref{fig:Fields}a shows the obtained \textit{SCC} weight map in the study area. The damping value, $\epsilon$, was fixed to 255, and the smoothing parameter, $\eta$, was 950. 
			
			The first arrival P-wave traveltimes can be used to obtain the 3D crustal velocity structure of the Central Alborz region. This 3D model can provide an insight into the detailed crustal velocity structure of the Central Alborz to better understand the fine-scale tectonics and seismic data transfer speed. Based on the event’s epicenter distribution and our study area, the 3D velocity model obtained by Afra et al.~\cite{afra2021three} can be an appropriate model. This model has been calculated using an iterative linearized, damped least-squares widely used inversion code \textit{SIMULPS}~\cite{evans1994user}. Because the recovered anomalies of the model have been confirmed by the two different (including seismology and gravity) geophysical methods. Moreover, the reliability of this velocity model was performed by different resolution tests which are including the checkerboard resolution (both dense and sparse for checking lateral and smearing resolutions, respectively), input initial model uncertainties, events location uncertainties, and Resolution Diagonal Element~\cite{afra2021three}. In this study, the region with the fairly resolution is surrounded by a black thick border according to ~\cite{afra2021three}. FIG.~\ref{fig:Fields}c represents this $3D$ velocity model from subsurface to depth of $30$ $km$. It should be noted that the $2D$ \textit{SCC} weight was also parameterized as the depth of the velocity model as shown in FIG.~\ref{fig:Fields}b.
			
			The coda and body wave quality factor (Q-factor) is a powerful tool to study thermal, compositional, and deformational characteristics of Earth's interior~\cite{singh2015study} in seismology. Generally, body or surface waves are attenuated with rates greater than the calculated rates for geometrical spreading. The inverse of the Q-factor, $Q^{-1}$, can separate the scattering, $Q^{-1}_{Sc}$, attenuation model (because of inhomogeneities within the earth) from intrinsic, $Q^{-1}_{i}$, attenuation (because of the geometrical spreading) along the propagation path of a seismic wave, so that $Q^{-1} = Q^{-1}_{Sc} + Q^{-1}_{i}$~\cite{wennerberg1993multiple}. Naghavi et al.~\cite{naghavi2012lg} obtained the Q-factor model for the Central Alborz. This model has been calculated by the Lg coda method using 1020 waveforms of the vertical component of 205 earthquakes with $3.5\leq \textbf{M}_L \leq 6.5$ recorded by 35 short-period seismic stations between 2000-2009. In this study, we used the Q-factor model calculated by Naghavi et al.~\cite{naghavi2012lg} for predicting source-receiver attenuation as depicted in FIG.~\ref{fig:Fields}d. 
	}}
	\begin{figure*}
		\centerline{\includegraphics[scale=0.8]{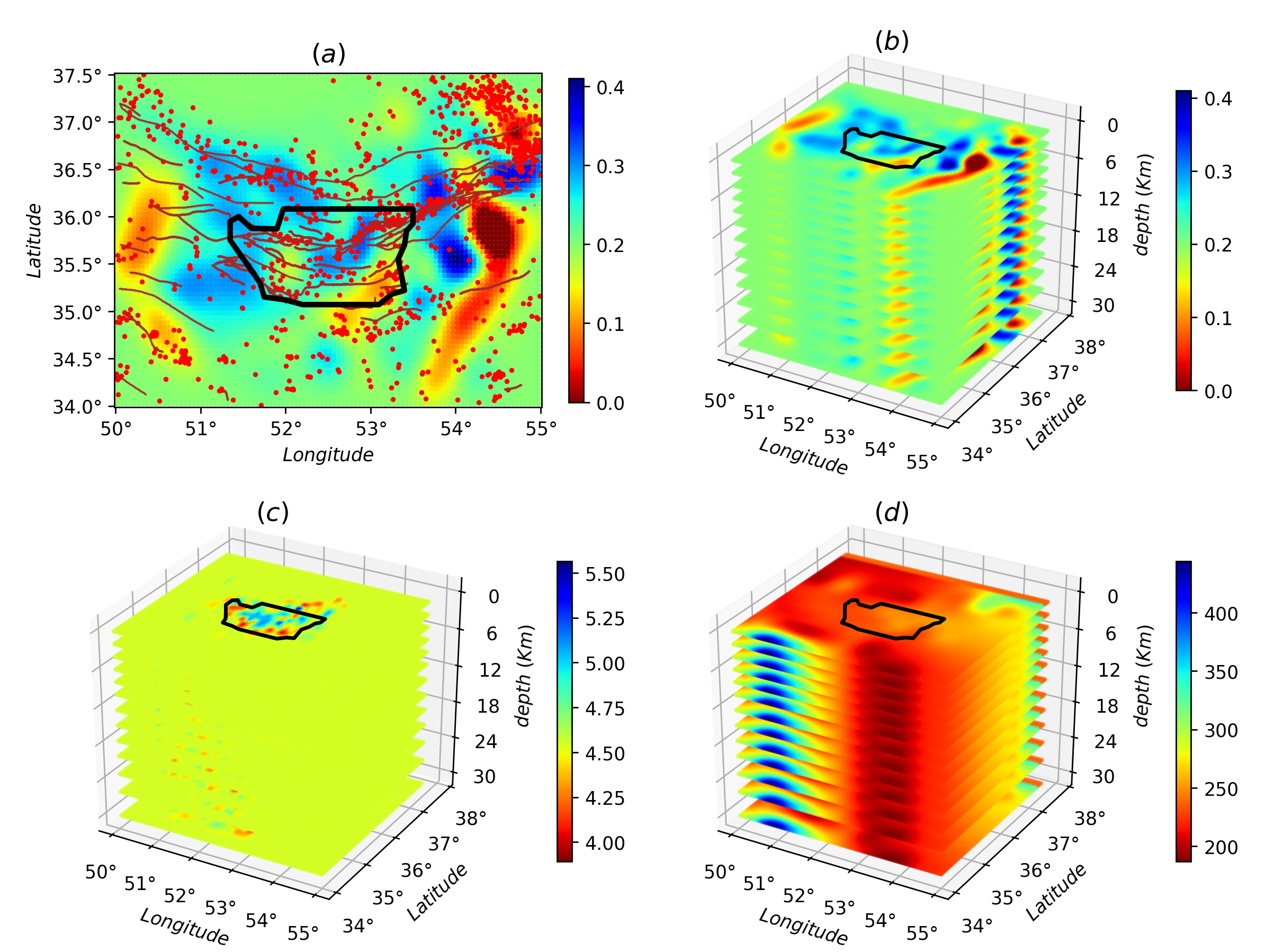}}
		\caption{(Color Online) (a) A two-dimensional (2D) projection of BTW lattice on the $x_1-x_2$ plane ($x_3=0$). The red lines show known faults, and red circles are the events with the magnitude of \textbf{M} $\geq 2.5$. The three-dimensional (3D) projection is shown for (b) Cross-Correlation Weight, (c) Q-factor~\cite{naghavi2012lg}, (d) P-wave velocity model~\cite{afra2021three} wherein The Cross-Correlation Weight and Q-factor are robust against $x_3$.}
		\label{fig:Fields}
	\end{figure*}
	
	\section{THE Dynamical MODEL}\label{SEC:Model}
	
	Our phenomenological dynamical model is a variant of dissipative continuous sandpile model~\cite{Lubeck,bak1987self}, that is implemented on top of a three-dimensional cubic lattice with coordination number $z=6$. The spatial extent of the region under study is the map presented in Fig.~\ref{fig:Phenomenona} and Fig.~\ref{fig:Fields}. We parameterized (meshed) the system so that the lattice points are fitted to existing data from which we picked the earth factors like the $q$-factor and the velocity filed. The resulting lattice consists of $N_1\times N_2=100\times 70$ nodes in each plate parallel to the earth surface, and totally $N_3=13$ horizontally plates are considered in the perpendicular direction as shown if Fig.~\ref{fig:Fields}, so that the lattice includes $N=L_1L_2L_3$ sites. To each site of the lattice $i$, we attribute three intrinsic quantities obtained from the real data of the earth explained in the previous section~\cite{afra2021three,naghavi2012lg}: the weight function ($W_i$), the quality factor ($Q_i$), and the velocity Model ($V_i$). The weight and other functions have designed so that their values for the connection between two neighboring sites is $f_{ij}=\frac{1}{2} (f_i + f_j)$, wherein $f_{ij}$ can be $W_{ij}$, $Q_{ij}$, and $V_{ij}$. The resulting weight field, the $Q$-factor, and the velocity field model are shown in Figs.~\ref{fig:Fields}b, c and d. Using the weight field we construct the lattice as represented in Fig.~\ref{fig:Fields}a, where the position of faults and the points at which earthquakes have taken place are shown. The threshold field to be used in the dynamics of the system is proportional to the field given in Fig.~\ref{fig:Fields}b, i.e. $\epsilon_i^{\text{th}}=\sum W_{ij}$. Once the network is constructed, we define the following dynamics, which is based on the local energy/stress in each site $i$, denoted by $\epsilon_i$, taking values in the range $[0,\epsilon_i^{\text{th}}\equiv \sum_{j=1}^{z} W_{ij}]$ (the summation $j$ is over neighbors of $i$), so that a local status of the system is identified by the set $\left\lbrace \epsilon_i \right\rbrace_{i=1}^{N}$. The initial state of the system is chosen to be random with uniform distribution. \\
	
\begin{figure}
		\centerline{\includegraphics[scale=0.6]{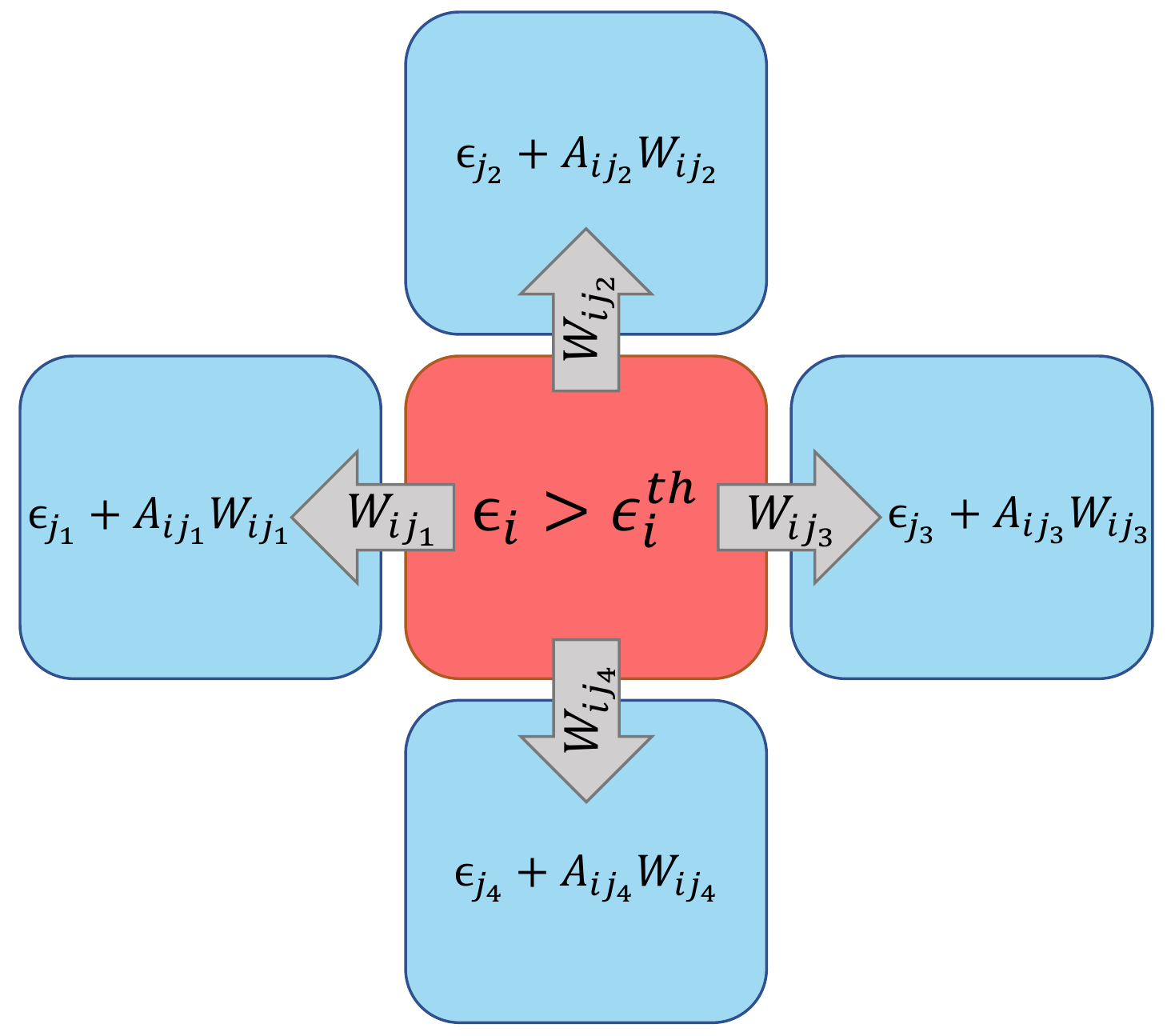}}
		\caption{A schematic 2D set-up of the sandpile model. The rad site, $i$, is unstable since $\epsilon_i^{th}>\epsilon_i$ and sends $W_{ij}$ stress to any neighboring sites, but the neighboring sites get $A_{ij}W_{ij}$ energy.}
		\label{fig:schematic}
	\end{figure}

	The dynamics of the system are defined by local relaxations generated by local slipping of the fault, i.e. distributing the stress excess through the neighboring regions. The rate of stress transfer is related to the $Q$-factor, velocity model, and the weight of the connections. The local stimulation of a region is external and is implemented on a randomly chosen site $i$ via $\epsilon_i\rightarrow \epsilon_i+r$, where $r$ is a flat random number between 0 and 1), which favors a local slip of this site. This site is however static if its accumulated stress is lower than a threshold $\epsilon^{\text{th}}_i$ as a consequence of local static friction~\cite{??}. If $\epsilon_i$ exceeds $\epsilon_i^{th}$, then site $i$ is called unstable, leading to a toppling process (local relaxation), during which $\epsilon_i \to \epsilon_i - \Delta_{ij}$, where matrix $\Delta$ is defined as
	\begin{equation}
		\begin{split}
			\Delta_{i,j}=\left\lbrace \begin{matrix}
				-w_{ij} A_{ij} &\ \ \  \text{if}\ i \ \text{and}\ j \ \text{are neighbors}\\
				\\
				\epsilon_i^{th} & \text{if}\ i=j\\
				\\
				0 & \text{otherwise}
			\end{matrix}\right.  ,
		\end{split}
	\end{equation}
	where $A_{ij}$ is the inelastic attenuation factor~\cite{singh2015study}
	\begin{equation}\label{eq:threshold}
		A_{ij}(f,r)= \dfrac{A_0}{\sqrt{r_{ij}}} e^{-\pi f \dfrac{t_{ij}}{Q_{ij}}},
	\end{equation}
	where $A_0$ is an amplitude, $f$ is frequency, $r_{ij}$ and $t_{ij}\equiv \frac{r_{ij}}{V_{ij}}$ are the distance and the travel time between sites $i$ and $j$. In our coarse grained model, we ignore the dependence on the frequency by setting $f=const.$. Moreover, to recover conservative dynamics in the limit $Q\rightarrow\infty$, we set $A_0=\sqrt{r_{ij}}$, so that the inelastic attenuation factor reduces to $A_{ij}=\exp -r_{ij}/Q_{ij}V_{ij}$. The toppling rule is schematically shown in Fig.~\ref{fig:schematic}. Also we have $r_{ij}=4.52$\si{\km} in the $x_1$ direction, $r_{ij}=5.55$\si{\km} in the $x_2$ direction, and $r_{ij}$ varies from $2$\si{\km} to $4$\si{\km} in the $x_3$ direction.\\

	Burst dynamics is a popular property of sandpiles, manifested by prominent avalanches which occur with a low frequency, called sometimes rare events~\cite{najafi2012avalanche,lubeck2000moment,najafi2016bak,cheraghalizadeh2017mapping,najafi2020geometry,najafi2018sandpile}. Avalanches in our model are defined as a chain of local relaxations (topplings) occurring as a consequence of an external stimulus. In fact, when one site becomes unstable by an external stimulus, it topples, and as a consequence, the neighboring site might become unstable and topple in their turn, so that a chain of relaxations take place up to a time where no further unstable site is found. The duration $D$ and the size $S$ of avalanche are defined as the lifetime and the total number of topplings in the avalanche respectively. Then another random site for stimulation is chosen and so on. The average local stress grows almost linearly with time until reaching a stationary state after which the amount of stress that leaves the system through the boundaries is statistically equal to the number of input stress, for a good review see~\cite{najafi2021some}. 
	
	\section{measures and results}\label{NUMDet}
\begin{figure*}
		\centerline{\includegraphics[scale=0.7]{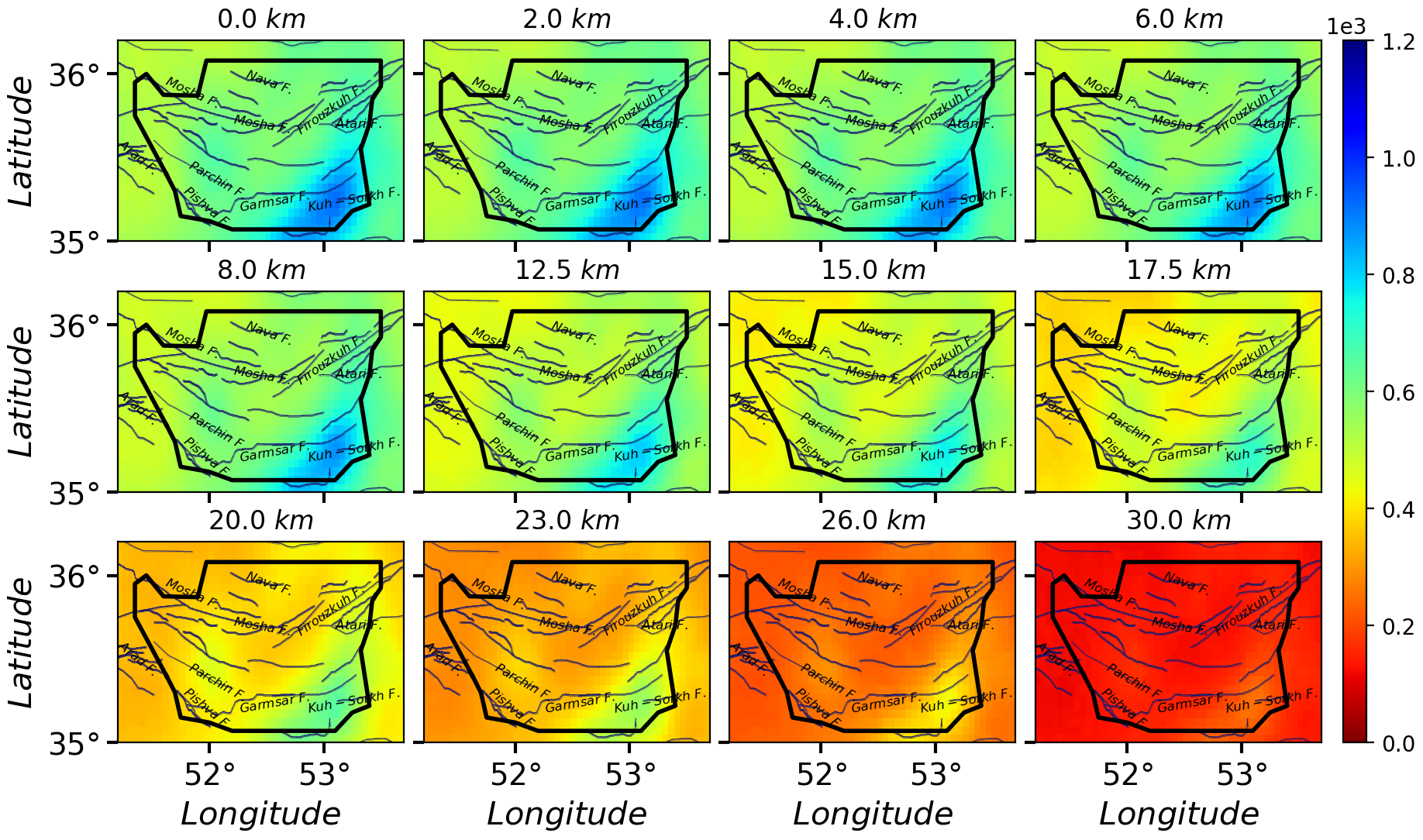}}
		\caption{The activity Field for $1000000$ avalanches in the whole space, wherein each avalanche starts from a random site. The solid lines and black border indicate the known faults and the region with good resolution, respectively.}
		\label{fig:Activity_Field_Random}
	\end{figure*}

\begin{figure*}
		\centerline{\includegraphics[scale=0.7]{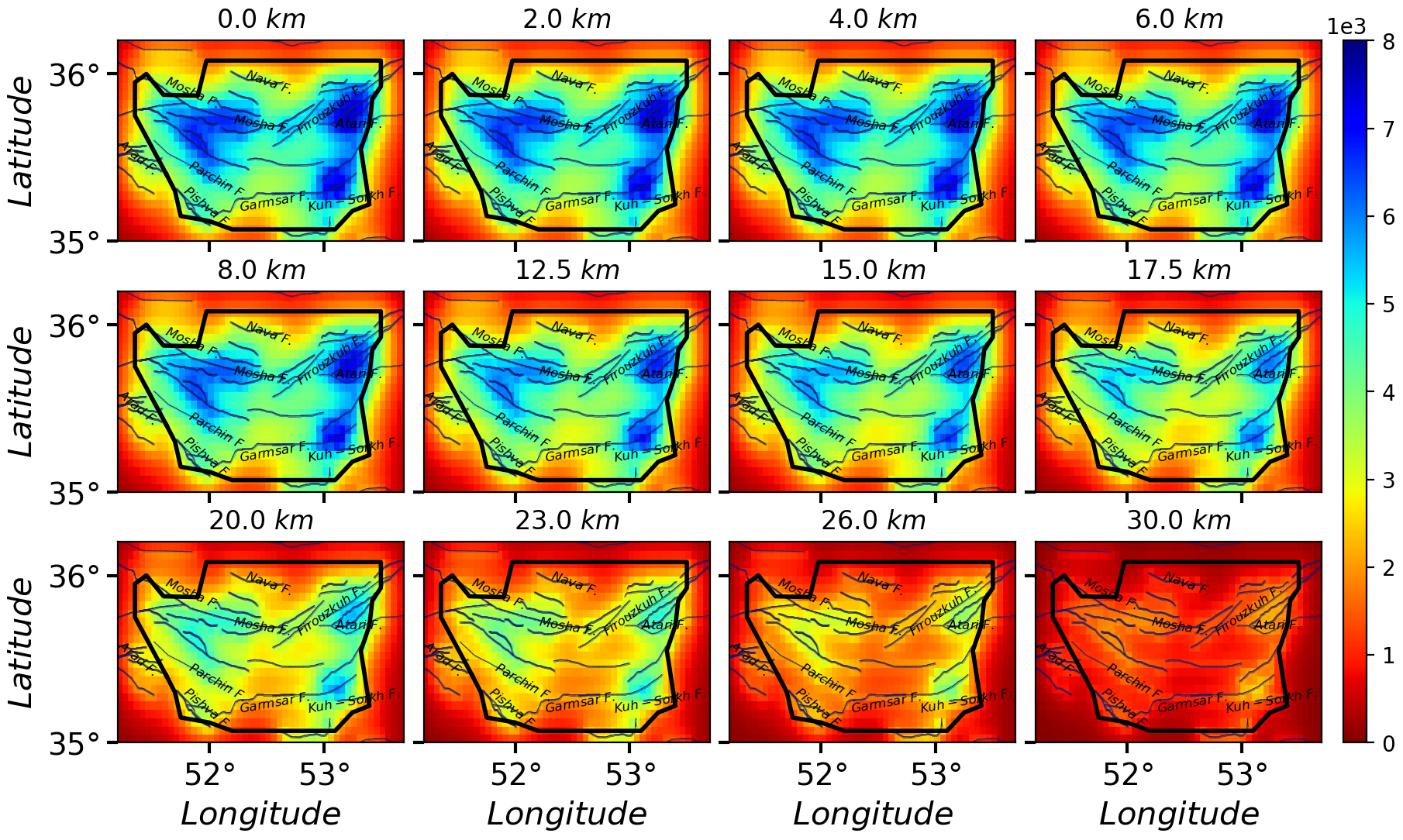}}
		\caption{The activity Field for $1000000$ avalanches in the whole space, wherein each avalanche starts from a Fault site. The solid lines and black border represent the known faults and the region with good resolution, respectively.}
		\label{fig:Activity_Field_Faults}
	\end{figure*}

\begin{figure*}
		\centerline{\includegraphics[scale=0.7]{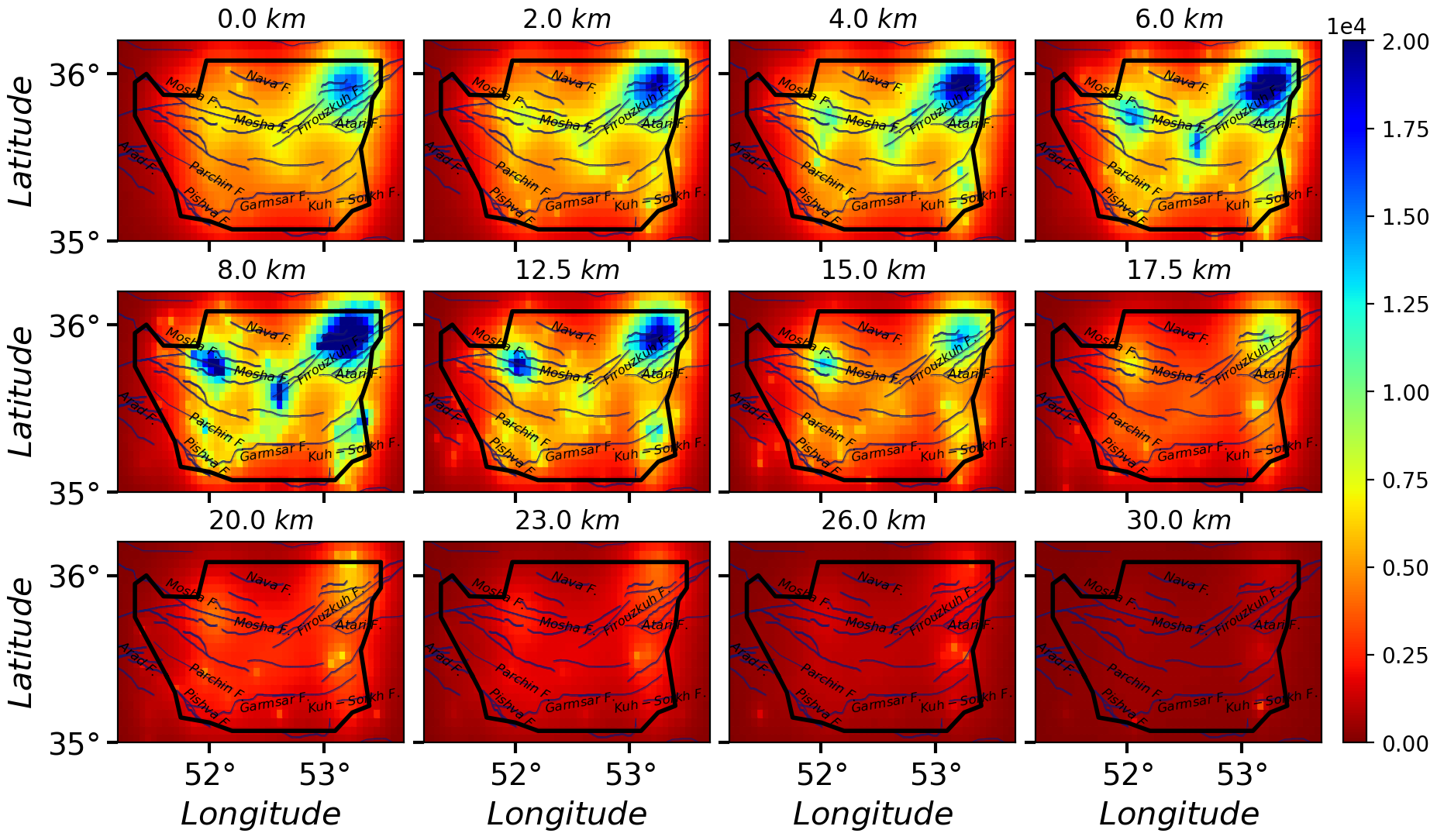}}
		\caption{The activity Field for $1000000$ avalanches in the whole space, wherein each avalanche starts from an event \textbf{M} $\geq 2.5$ site. The solid lines and black border indicate the known faults and the region with good resolution, respectively.}
		\label{fig:Activity_Field_events}
	\end{figure*}

\begin{figure*}
		\centerline{\includegraphics[scale=0.7]{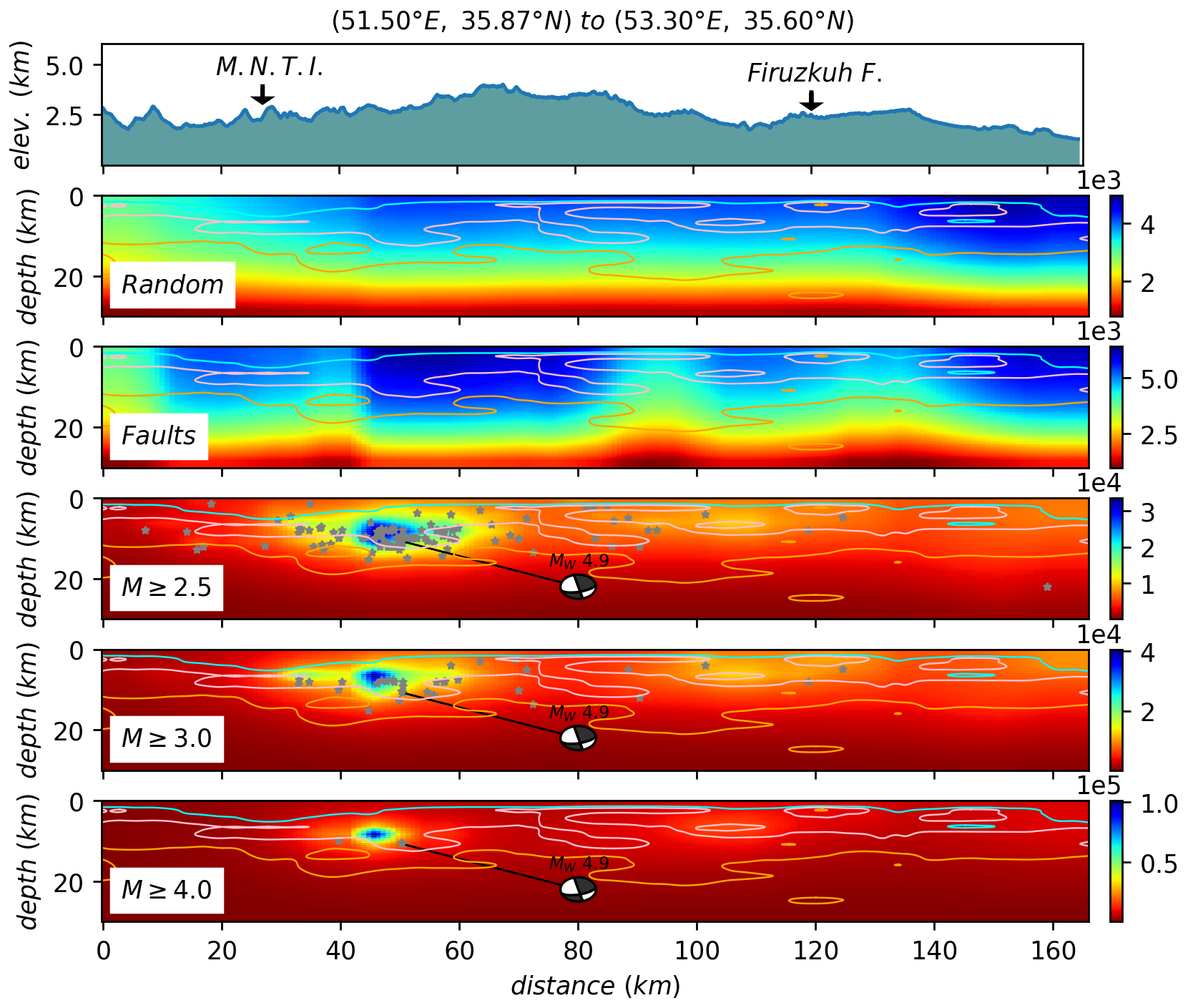}}
		\caption{An activity field for the cross-section along (51.50$^{\circ}$E, 35.87$^{\circ}$N) to (53.30$^{\circ}$E, 35.60$^{\circ}$N) for the dynamics with avalanches in the whole space. All hypocenters of earthquakes within $\pm$7 km distance from the profile are projected. The focal mechanism solution show an earthquake with \textbf{M}$_W$ 4.9 occurred at (52.05$^{\circ}$E, 35.78$^{\circ}$N, 7km) on May 7$^{th}$, 2020 at 20:18:21.00 which was reported by IrSC. The solid cyan, pink, and orange contours represent the P-wave velocity of 5.8, 6.1, 6.3 kms$^{-1}$, respectively, calculated in Ref.~\cite{afra2021three}. The M.N.T.I abbreviation refers to the Mosha-North  Tehran  fault  intersection (see FIG. 1).}
		\label{fig:cross-section}
	\end{figure*}

	In this section, we present the results of the simulation of our phenomenological model. The size of the lattice is fixed as explained in the previous section. We have tested five kinds of external stimuli, which are listed bellow: \\
	\begin{itemize}
		\item $\mathbb{I}$: completely random stimuli (the sites for external stimuli is completely random chosen),
		\item $\mathbb{II}$: fault stimuli (the sites for external stimuli are randomly chosen on the faults),
		\item $\mathbb{III}$: stimulation of Highly active regions, (the sites for the external stimuli are randomly chosen from sites of events with amplitudes of \textbf{M} $\geq 2.5$),
		\item $\mathbb{IV}$: stimulation of medium active regions (the sites for the external stimuli are randomly chosen from sites of events with amplitudes of \textbf{M} $\geq 3.0$),
		\item $\mathbb{V}$: stimulation of Highly active regions (the sites for the external stimuli are randomly chosen from sites of events with amplitudes of \textbf{M} $\geq 4.0$).
	\end{itemize}
	The case $\mathbb{I}$ represents the case were the local stress excess (resulting to an earthquake) takes place in a completely random region, while the case $\mathbb{II}$ realizes the situations where the earthquake starts on the fault. The cases $\mathbb{III}$, $\mathbb{IV}$, and $\mathbb{V}$ capture the cases where its starts from more active sites which is classified to \textbf{M} $\geq 2.5$, \textbf{M} $\geq 3.0$ and \textbf{M} $\geq4.0$. For each case, we have generated over $10^6$ samples, i.e. the avalanches in the stationary states. The activity field for $10^6$ samples for the cases $\mathbb{I}$, $\mathbb{II}$ and $\mathbb{III}$ are shown in Figs.~\ref{fig:Activity_Field_Random},~\ref{fig:Activity_Field_Faults}, and~\ref{fig:Activity_Field_events} respectively. By looking at Fig.~\ref{fig:Fields}a, we see a good correlation between the average activity field and the weight field. A much similarity is observed between the case $\mathbb{I}$ and the weight field but as the pattern of stimuli changes (cases $\mathbb{II}$, $\mathbb{III}$, $\mathbb{IV}$, and $\mathbb{V}$), the activity field show different patterns. Especially for the fault stimuli (case $\mathbb{II}$) the activity in the vicinity of the fault positions is much higher, which is rather expected. A more realistic situation is the case where the stimulation takes place in the vicinity of the highly active points (cases $\mathbb{III}$, $\mathbb{IV}$, and $\mathbb{V}$), e.g. Fig.~\ref{fig:Activity_Field_events} since these points are more active with respect to the rest regions. Apart from the observational data (the points where the earthquake has taken place), this activity field predicts the activity of the other regions, and explicitly shows important points which can be potentially the starting point of the upcoming earthquakes. Figure~\ref{fig:cross-section} shows a vertical cross-section of  belt passing the Mosha fault, extended from along (51.50$^{\circ}$E, 35.87$^{\circ}$N) to (53.30$^{\circ}$E, 35.60$^{\circ}$N), where M.N.T.I refer to the Mosha-North Tehran fault intersection. Interestingly this shows an earthquake with \textbf{M}$_W$ 4.9 occurred at (52.05$^{\circ}$E, 35.78$^{\circ}$N, 7km) on May 7$^{th}$, 2020 (the blue region in the three lower graphs). This shows that our model is surprizingly working, i.e. it nicely shows the active regions that were active in past.\\
	
	Although we applied the random event location (results depicted in FIG.~\ref{fig:Activity_Field_Random}), the potential to produce an earthquake mostly can be appeared around Kuh-Sorkh and Parchin faults. By exciting cells on the known faults, this potential mostly appears around Atari-Firuzkuh, western of Mosha, Kuh-Sorkh, and Parchin faults. This potential can become apparent for cells excited by events' location around Atari-Firuzkuh, Firuzkuh, and western of Mosha. A simple comparison for these activities (FIG.~\ref{fig:Activity_Field_Random},~\ref{fig:Activity_Field_Faults}, and~\ref{fig:Activity_Field_events}) indicate the potential to produce an earthquake for superficial layers can be expected for micro-earthquakes (\textbf{M} $\leq$ $3.0$) which is in agreement with seismicity in the study area. But, this potential can produce a larger earthquake at the greater depths ($15$ to $30$ $km$) as experienced by historical earthquakes (see FIG.~\ref{fig:Phenomenona}). Our results are consistent with previous studies (e.g.~\cite{afra2021three}) that P-wave tomography and gravity inversion models reveal a region at depth ranges of $12.5$-$17.5$ $km$ (follow the orange contour in FIG~\ref{fig:cross-section} between distances of 60-90 km) for the potential of producing an earthquake with \textbf{M} $6.5$.
	
	The left-lateral Mosha fault, which is the most important internal fault to the central Alborz Mountains, can release a part of energy in the amount of the $\sim 4\ mmyr^{-1}$ range-parallel strike-slip motion. Vertical intersection distributions of the final activities models for Mosha fault are shown in Fig.~\ref{fig:cross-section}. This profile along ($51.50^{\circ}$E, $35.87^{\circ}$N) to ($53.30^{\circ}$E, $35.60^{\circ}$N) is $\sim 165\ km$ long. The random activity along this profile can clearly represent the upper crust (a thick layer up to $ 20\ km$ depth) in which most earthquakes can occur. This thickness is in agreement with the report by Abbassi et al.~\cite{abbassi2010crustal} obtained by the employment of a local seismic network. For the fault activity, the interface topography between the upper and middle crust improves realistically, which is in agreement with the bottom of event depths as reported in~\cite{tatar2012microseismicity}. Inspection of this intersection reveals that high activities seem to be around the Firuzkuh, Atari faults, and also Mosha-North Tehran fault intersection (M.N.T.I) in an expectation area (see~\cite{afra2021three}) which have a potential of the future earthquake with \textbf{M} $\sim 6$ to $6.5$. As shown in activities of events profiles (\ref{fig:cross-section}), two areas (distance $\sim 40-60\ km$ , $\sim 100-120\ km$) clearly highlight most likely parts on the Mosha fault which can experience earthquakes larger than \textbf{M} $\geq$ 4. These anomalies can be consistent with the occurred events. An earthquake (52.05$^{\circ}$E, 35.78$^{\circ}$N, 7km) with \textbf{M} 4.9 has occurred on the Mosha fault around the resolved activity anomaly (distances $\sim 40-60\ km$) on May 7$^{th}$, 2020 at 20:18:21.
	
\begin{figure*}
		\centerline{\includegraphics[scale=0.6]{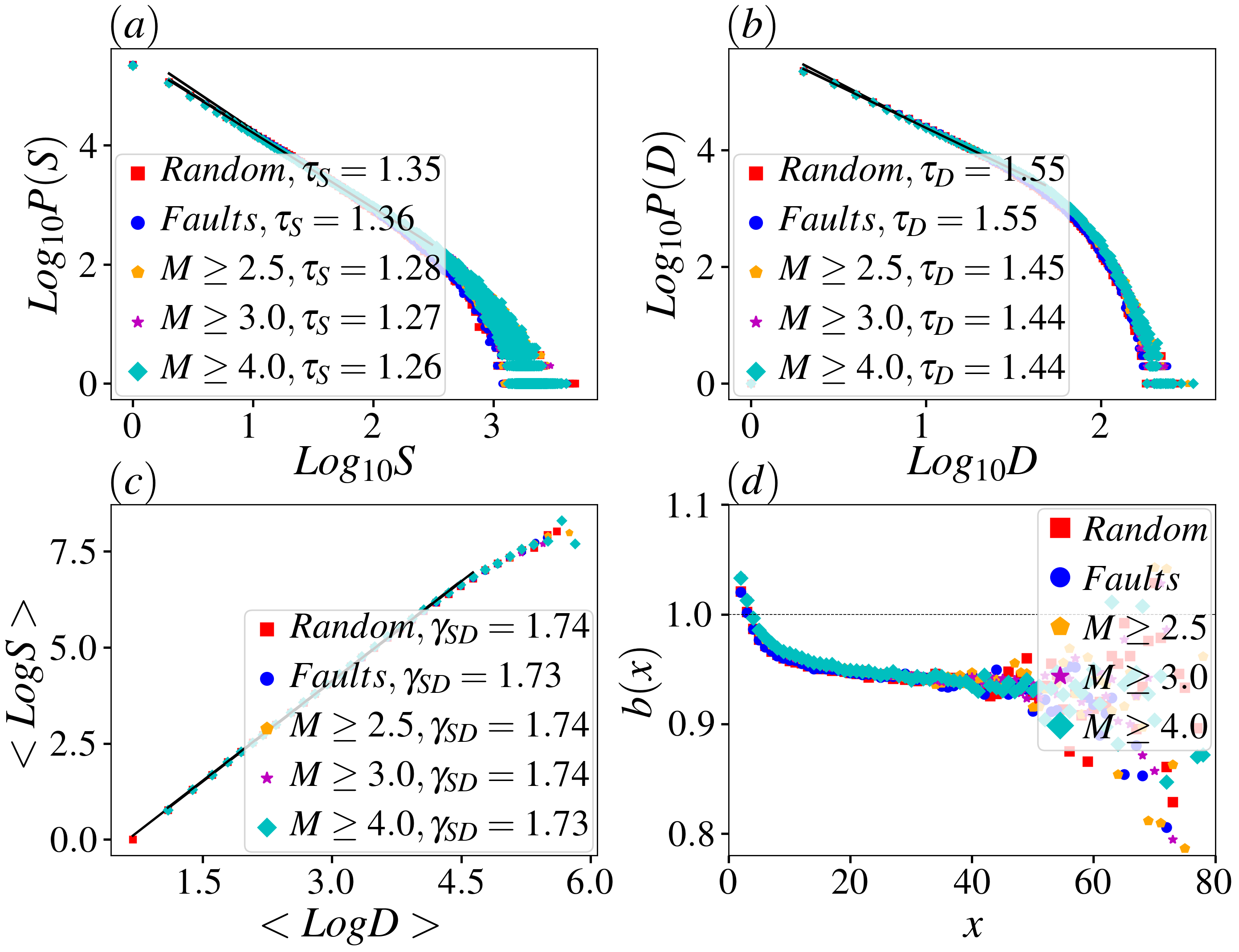}}
		\caption{The distribution function of (a) avalanche size $S$, (b) avalanche duration $D$. (c) The $\log$-$\log$ plot of $S$-$D$ scaling relation. (d) Activity dependent branching ratio $b(x)$ for instantaneous avalanche size $x$. All plots are for the dynamics with avalanches in the whole space}
		\label{fig:av_statisticsWithExt}
	\end{figure*}

	The results that are shown in the above figures are graphical demonstration of the situation that the region have. It is now worthy to quantify the universal behaviors of the model for the five cases that we introduced. In Fig.~\ref{fig:av_statisticsWithExt}, the distribution functions ($P$) of the avalanche duration $D$ and size $S$ are exhibited. In Fig.~\ref{fig:av_statisticsWithExt}a and b, we show the log-log plot of $P(S)$ and $P(D)$ respectively where the linear decrease in the signature of power-law decay $P(x)\propto x^{-\tau_x}$, $x=S,D$. Interestingly the exponents depend of the taken situation, see TABLE~\ref{TAB:exponents}. Noting that for the three-dimensional Bak-Tang-Weisenfeld (BTW) sandpile model, we have $\tau_S\simeq \frac{4}{3}$~\cite{lubeck2000moment,najafi2020geometry}, we see that for all cases, the exponents $\tau_S$ and $\tau_D$ coincide well with 3D sandpile. The scaling dimension $\gamma_{SD}$ defined by $S\propto D^{\gamma}$ is almost robust, i.e. $\gamma_{SD}=1.75\pm 0.008$, $1.73\pm 0.008$, $1.76\pm 0.009$, $1.74\pm 0.008$ and $1.74\pm 0.008$ for the cases $\mathbb{I}$ to $\mathbb{V}$ respectively. 
	
	To test the criticality of the system, one can use the {branching ratio} function defined by the conditional expected value $b(x)\equiv \mathbb{E}\left[\frac{S_{t+1}}{x} |S_{t}=x\right]$, where $\mathbb{E}$ is expected value. For the critical systems $\lim_{x\rightarrow 0}b(x)=1$ or is in the vicinity of unity~\cite{martin2010activity,alstrom1988mean,rahimi2021role,najafi2020avalanches,najafi2019effect}. This function has been shown in Fig.~\ref{fig:av_statisticsWithExt}d, for which $1<\lim_{x\rightarrow 0}b(x)<1.03$, showing that although not exactly (super-critical), but we are in pretty in the vicinity of the critical region, which is confirmed by other power-law behaviors.
	
	\section{Multi-fractal analysis}\label{SEC:MFA}
	Here we map the system to a time series to analyze the spectrum of the Hurst exponent. For an uncorrelated time series the Hurst exponent is a single value $H=0.5$~\cite{mandelbrot1985self,qian2004hurst}. For a mono-fractal time series, the Hurst exponent, $H$, is defined in terms of the asymptotic behaviour of the rescaled range. Consider a general time series $\left\lbrace I(t)\right\rbrace_{t=1}^n$, and the cumulative deviate (profile) series $I^*(t)\equiv \sum_{t'=1}^t \left(  I(t')-\bar I\right)$, where $\bar{I}\equiv \frac{1}{n}\sum_{t=1}^nI(t)$. The range is then defined by 
	\begin{equation}
		R(n)\equiv \text{max}\left\lbrace I^*(t)\right\rbrace_{t=1}^n-\text{min}\left\lbrace I^*(t)\right\rbrace_{t=1}^n
	\end{equation}
	The standard deviation is also defined as
	\begin{equation}
		S(n)\equiv \sqrt{\frac{1}{n}\sum_{t=1}^n\left( I(t)-\bar{I}\right)^2}.
	\end{equation}
	Having defined these parameters, one can obtain the Hurst exponent $H$ using the relation
	\begin{equation}
		\mathbb{E}\left[\frac{R(n)}{S(n)}\right]\propto n^H.
	\end{equation}
	This is applied for the case where there is an ensemble set. For the case where we have a single time series, we divide the time series into non-overlapping segments $ \lfloor M_{s}=M/s \rfloor $ with the equal lengths $s$. Then the generalized variance is defined by the relation
	\begin{equation}
		G_{\mathrm{w}}(s)={\sum_{v=1}^{M_s} \vert I^*(vs)-I^*((v-1)s)\vert^{\mathrm{w}}}.
		\label{eq:Ftau}
	\end{equation}
	This function shows often power-law behavior with $s$ like $G_{\mathrm{w}}(s) \sim s^{wH-1}$~\cite{kantelhardt2002multifractal,arias2021testing}.\\
	
\begin{figure*}
		\includegraphics[scale=0.5]{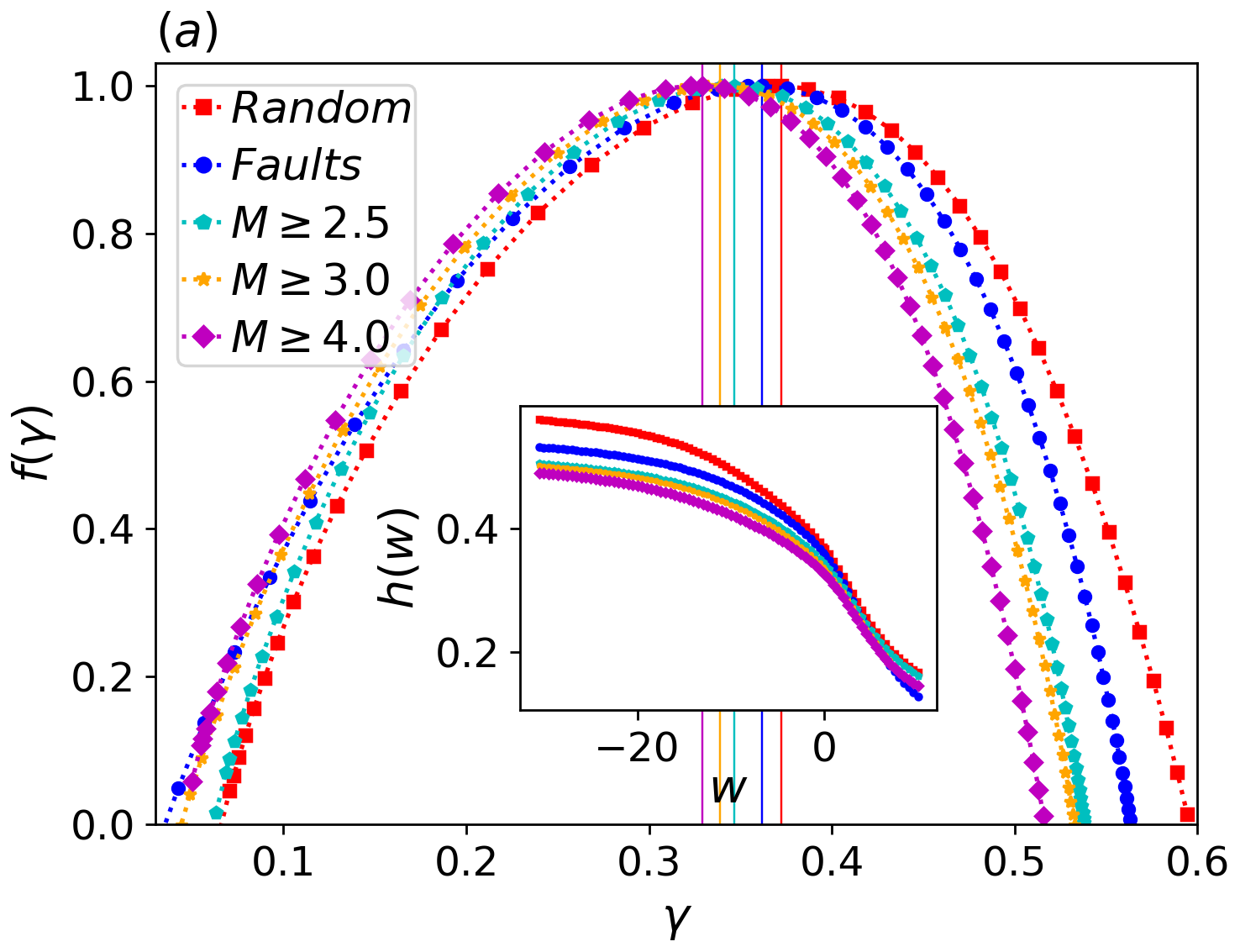}
		\includegraphics[scale=0.5]{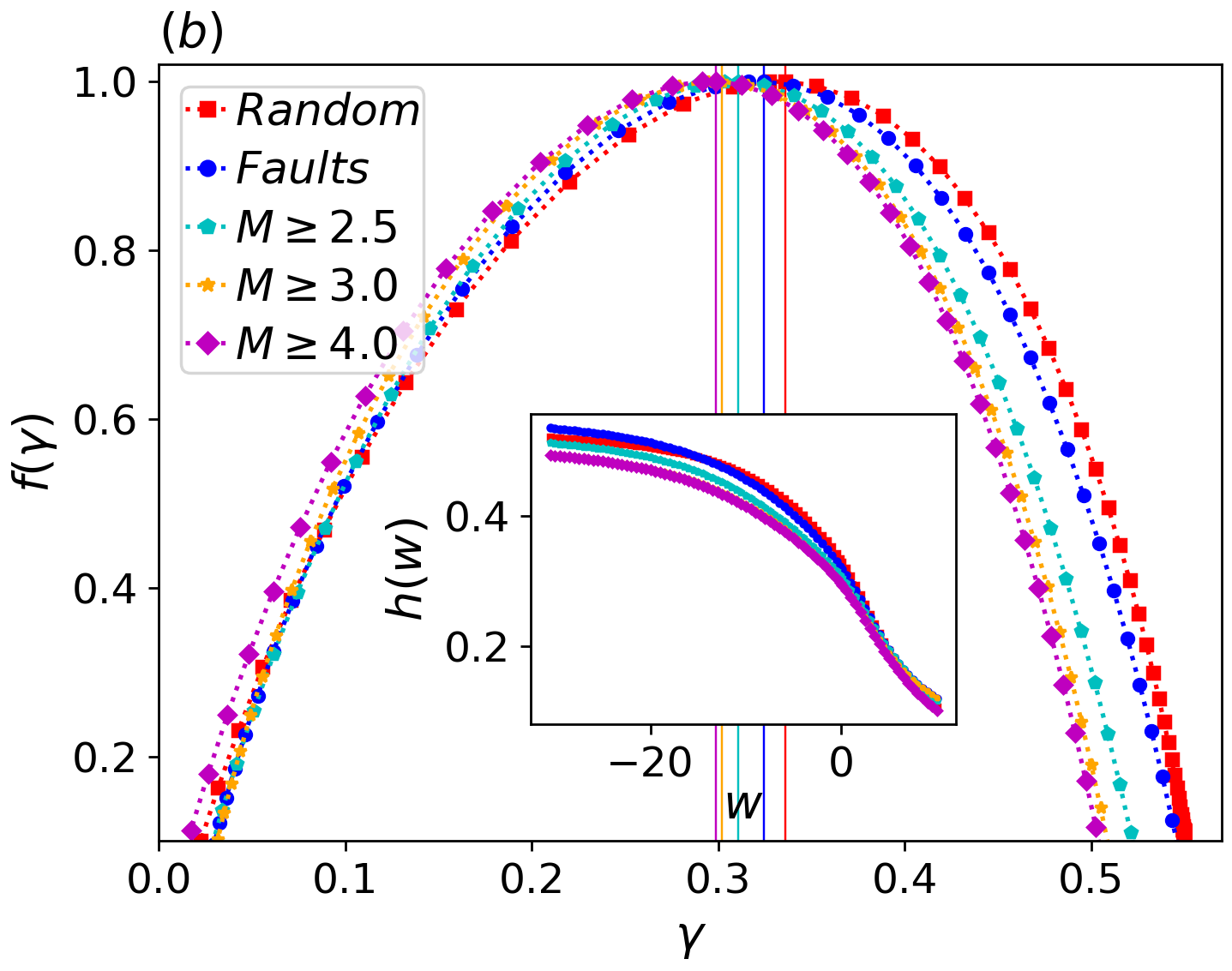}
		\caption{(a) Main panel: $f(\gamma) $ vs. $\gamma$ for various dynamics for the dynamics with avalanches in the whole space. inset: $h(w)$ vs.$  \mathrm{w} $ for various dynamics. (b) The same for the dynamics with avalanches inside the high-resolution box.}
		\label{fig:gamma}
	\end{figure*}

\begin{figure*}
		\includegraphics[scale=0.6]{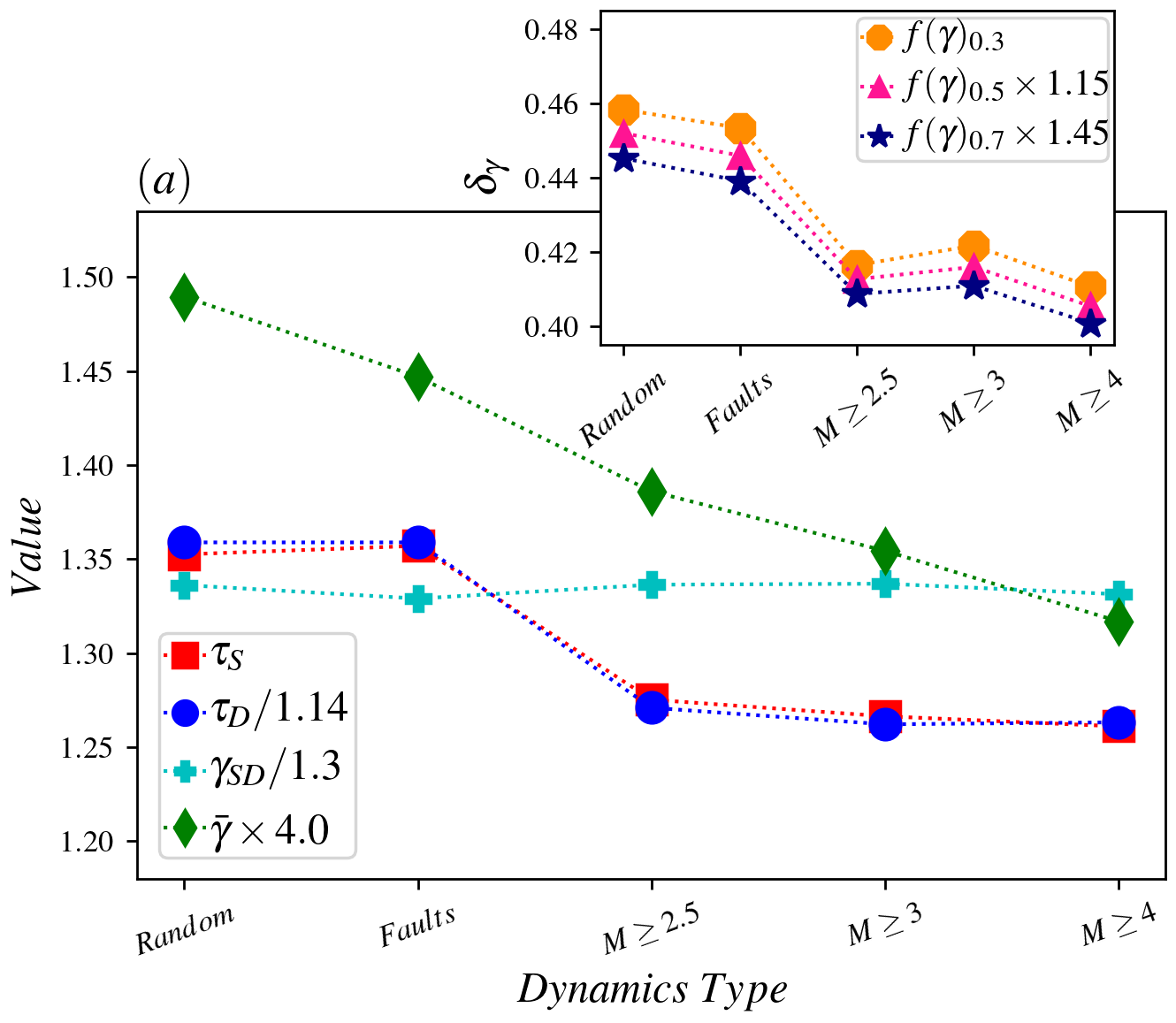}
		\includegraphics[scale=0.6]{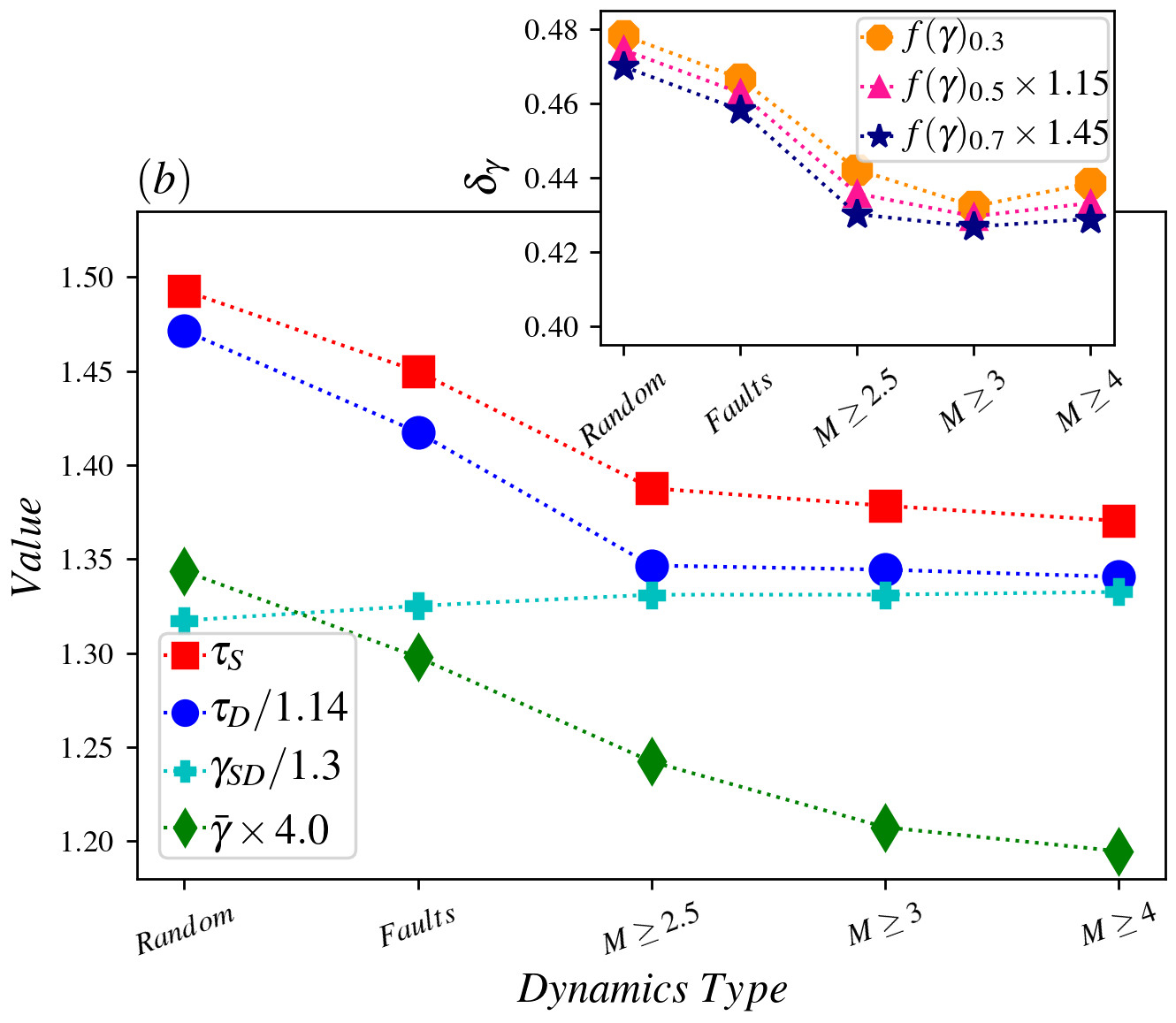}
		\caption{(a) Main panel: The value of different exponents vs. dynamics types for the dynamics with avalanches in the whole space. inset: The width of $f(\gamma)$ function in different values against various dynamics types. (b) The same for the dynamics with avalanches inside the high-resolution box.}
		\label{fig:exponents}
	\end{figure*}

	Now we consider the multifractal time series with a spectrum of the Hurst exponent (for the mono-fractal time series, the spectrum is peaked with a zero width). This spectrum lets us know the mean as well as the fluctuations of the Hurst exponent, helping us to distinguish the type of correlations (and anti-correlations) existing in the system. We define a variance for each of the segments $v=1,2,3....M_{s}$ by~\cite{kantelhardt2002multifractal}
	\begin{equation}\label{eq:MAk}
		F^{2}(v,s)=\frac{1}{s}{\sum_{i=1}^{s} (I^*\left((v-1)s+i\right) -\overline{I^*}(v))^{2}},
	\end{equation}
	where, $\overline{x}(v)$ represents the mean of $x$ over the segment $v$. The $\mathrm{w}^{\text{th}}$ moment is then obtained using
	\begin{equation}\label{eq:DFA3}
		F_\mathrm{w}(s)\equiv\left\lbrace \frac{1}{N_{s}}{\sum_{i=1}^{N{s}} [F^{2}(v,s)]^{\mathrm{w}/2}}\right\rbrace ^{1/\mathrm{w}}
	\end{equation}
	The typical behavior of $F_w(s)$ is as follows~\cite{arias2021testing}
	\begin{equation}\label{eq:Fw}
		F_{\mathrm{w}}(s) \sim s^{h(\mathrm{w})}. 
	\end{equation}
	where $h(\mathrm{w})$ is the corresponding exponent that is related to the Hurst exponent. To extract the spectrum of the Hurst exponent we use the standard multifractal analysis. Therefore, we use the generalized variance Eq.~\ref{eq:Ftau}, but now with a new exponent $G_{\mathrm{w}}(s) \sim s^{\tau(\mathrm{w})}$
	where $\tau(\mathrm{w})$ represents the classical multifractal scaling exponent. This exponent is related to $h(\mathrm{w})$ for stationary and normalized time series,
	\begin{equation}\label{eq:tauq_H}
		\tau(\mathrm{w})=\mathrm{w}h(\mathrm{w})-1.
	\end{equation}
	The Legendre transform of the generalized scaling exponent $\tau(\mathrm{w})$ gives the multifractal function as follows
	\begin{equation}\label{eq:Legendre}
		\begin{split}
			f(\gamma)=\gamma \mathrm{w}-\tau(\mathrm{w})
		\end{split}
	\end{equation}
	where $\gamma=\dfrac{{\partial \tau(\mathrm{w})}}{{\partial \mathrm{w}}}$. Finally by a simple replacement one obtains \begin{equation}\label{eq:f_gamma}
		f(\gamma)=\mathrm{w}(\gamma-h(\mathrm{w})) +1,
	\end{equation}
	which gives the spectrum of the Hurst exponent. \\
	\begin{figure*}
		\centerline{\includegraphics[scale=0.8]{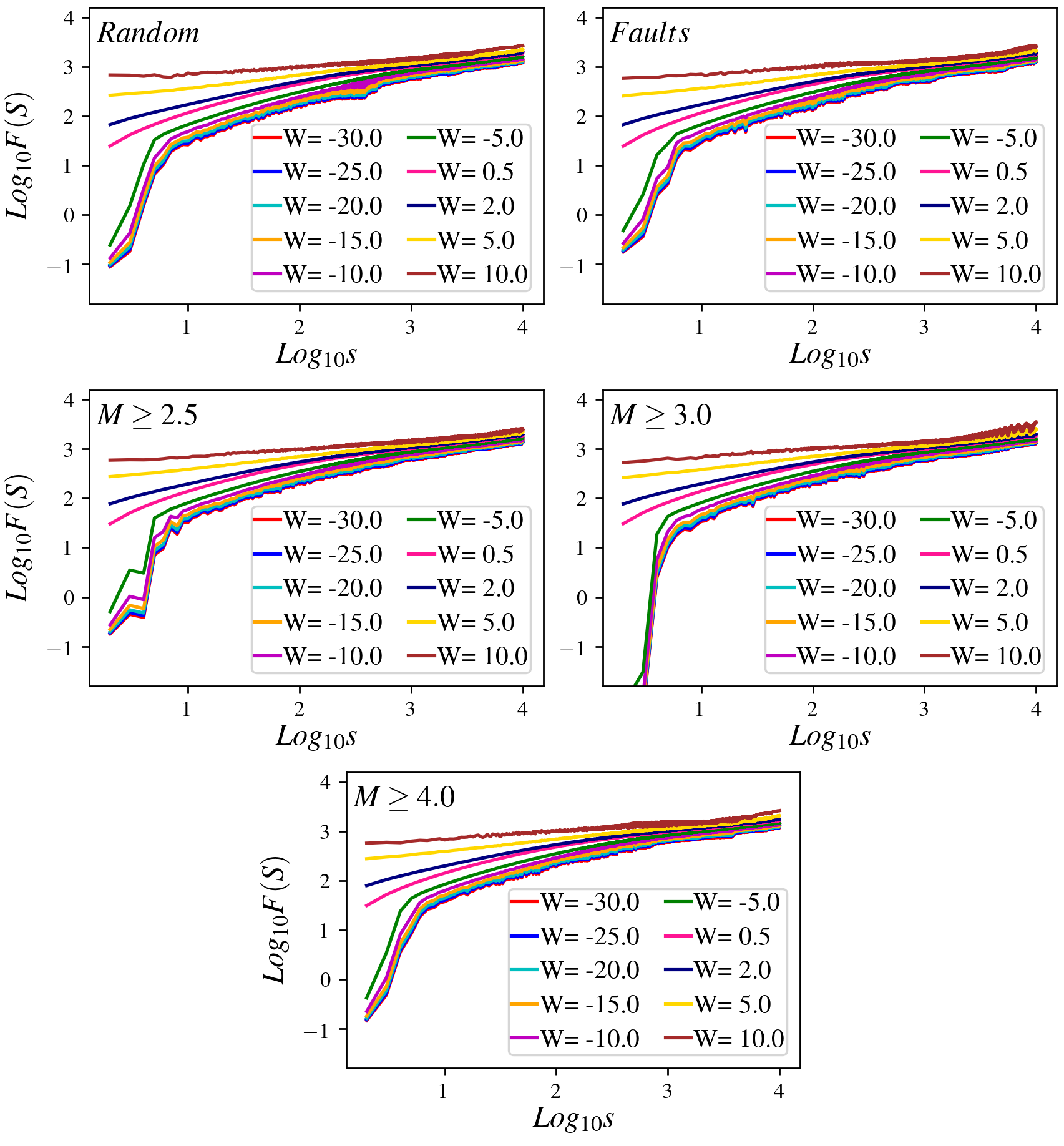}}
		\caption{Calculation of the statistical function $F_w$ using Eq.~(\ref{eq:DFA3}) for the dynamics with avalanches in the whole space. The function of $F_w$ vs $s$ display power laws $F_{w}(s) \sim s^{h(w)}$, where $h(w)$ depend on $w$. This feature demonstrates that the time series is a multifractal.}
		\label{fig:F}
	\end{figure*}
	The function $F(S)$ is shown in Fig.~\ref{fig:F} for the cases $\mathbb{I}$ till $\mathbb{V}$ and for various amounts of $\mathrm{w}$, where a power-law behavior is evident in a large interval (nearly two decades). The exponents of these graphs are $h(\mathrm{w})$. If $h(\mathrm{w})$ is the same for all $\mathrm{w}$ values, then we have mono fractal with Hurst exponent $h(\mathrm{w}=2)$. For this case $f(\gamma)$ would be a peaked function around $\gamma=h(\mathrm{w}=2)$ with zero width. Fig.~\ref{fig:gamma} shows that this is not the case, and we are facing with a strong multifractal time series for all cases. Fig.~\ref{fig:gamma}a shows the result for which we let the avalanches go beyond the almost square region identified in Fig.~\ref{fig:Phenomenona}b (high-resolution region), i.e. the region with high resolution that we are more confident about the weight field that we obtained. Fig.~\ref{fig:gamma}b shows the results for the case where we restrict the avalanches to the square. For both cases, we see that the width of $f$ is pretty high, and the peak position varies with the method of stimulation. Even for completely random stimulation, the peak is around $\bar{\gamma}=0.37$ ($0.36$) for the avalanches in the whole space (inside the high-resolution box), both being lower than $0.5$, showing that the system is anticorrelated. The exponents are shown in the table~\ref{TAB:exponents}.
	\begin{table*}
		\begin{tabular}{|c | c| c| c|c|c|}
			\hline   & case $\mathbb{I}$ (random)  &  case $\mathbb{II}$ (faults) & case $\mathbb{III}$ ($M\ge 2.5$) & case $\mathbb{IV}$ ($M\ge 3$) &  case $\mathbb{V}$ ($M\ge 4$)\\
			\hline $\bar{\gamma}$ & $0.37$ & $0.36$ & $0.35$ & $0.34$ & $0.33$  \\
			\hline $\delta\gamma$ & $0.46$ & $0.45$ & $0.42$ & $0.42$ & $0.41$  \\
			\hline $\tau_S$ & $1.35\pm 0.005$ & $1.36\pm 0.005$ & $1.28 \pm 0.004$ & $1.27 \pm 0.004$ & $1.26 \pm 0.004$  \\
			\hline $\tau_D$ & $1.55\pm 0.016$ & $1.55\pm 0.016$ & $1.45 \pm 0.01$ & $1.44 \pm 0.009$ & $1.44 \pm 0.009$  \\
			\hline
			\hline $\bar{\gamma}$ & $0.34$ & $0.32$ & $0.31$ & $0.30$ & $0.30$  \\
			\hline $\delta\gamma$ & $0.48$ & $0.47$ & $0.44$ & $0.43$ & $0.44$  \\
			\hline $\tau_S$ & $1.49\pm 0.008$ & $1.44\pm 0.008$ & $1.39 \pm 0.007$ & $1.38 \pm 0.007$ & $1.37 \pm 0.007$  \\
			\hline $\tau_D$ & $1.68\pm 0.025$ & $1.62\pm 0.024$ & $1.54 \pm 0.021$ & $1.53 \pm 0.022$ & $1.53 \pm 0.022$  \\
			\hline
		\end{tabular}
		\caption{Upper Table: The values of exponents  $\bar{\gamma}$, $\delta\gamma$, $\tau_S$, and $\tau_D$ for the dynamics with avalanches in the whole space, wherein the $\delta\gamma$ is defined as the width of $f(\gamma)$, which is the length of the interval between two successive $f(\gamma)=0.3$. Lower Table: The same for the dynamics with avalanches inside the high-resolution box.}
		\label{TAB:exponents}
	\end{table*}
	Using the fact that the exponent of autocorrelation function $\xi$ is related to the Hurst exponent like $\xi=2-2H$~\cite{ivanov2009levels} this result uncovers that the time series are anti-correlated, meaning that a large event is often followed by a small event and vice versa. This effect magnifies when the stimulation is more selective, i.e. for the fault stimuli, it is $\bar{\gamma}=0.36$, and for highly active stimuli case ($\mathbb{V}$) it is even smaller, $\bar{\gamma}=0.33$ for the case where avalanches are allowed to go through all the space. 
	The main results of our analysis are gathered in Fig.~\ref{fig:exponents}a and b (in whole space and limited to the high-resolution box respectively), where the relation between various exponents are shown in terms of the strategy of stimuli. The decrease of the Hurst exponent (green diamond symbols) with the stimuli strategy is evident in these graphs, reflecting the fact that the time series becomes more and more anti-correlated. Along with this, the exponents $\tau_S$ and $\tau_D$ decrease, but $\gamma_{SD}$ are almost robust. 
	
	\section{Conclusion}
	In this study, we focused on the earthquakes in central Alborz, Iran. In the first part of the paper, we explored the properties of the earth in the region under study, as well as the rate of earthquakes. It helped us to construct a phenomenological model which is much similar to the continuous dissipative sandpile model in which the energy dissipation is related to the quality factor and the velocity model of the earth. The weight function which was obtained using the signals-cross-correlation of the real seismic activities was used to estimate the weight field which was employed for distributing the energy to the neighboring sites in each toppling. Our model is based on external stimuli, the location of which can be (I) random, (II) on the faults, (III) on the highly active points in the region. The rate of earthquakes was shown to be related to the total activity field over the region of study.  Some universal behaviors of the system are shown to be related to the scheme taken for the initial stimuli. The second part of the paper was devoted to the Multi-fractal analysis, which is exploited to extract the spectrum of the Hurst exponent of time series. The time series for each scheme was analyzed separately by multifractal analysis, for all of which the average Hurst exponent is shown to be lower than $0.5$. This is an intrinsic property of anti-correlated time series, for which a large rare event is expected to be followed by a small event. The stimulation of highly active regions (in our study, the points with energies \textbf{M} $\geq4.0$), a lowest average Hurst exponent is obtained, meaning that we have the strongest anti-correlated system in this case. An overall phase diagram for the model is sketched for all schemes that were considered in this paper.
	
	\bibliography{refs}
	
\end{document}